\documentstyle[preprint,aps,epsf]{revtex}

\def\half{{\textstyle{1\over2}}}
\def\quar{{\textstyle{1\over4}}}
\def\six{{\textstyle{1\over3!}}} 
\def\Ii{{\textstyle{1\over i}}}
\def\beq{\begin{equation}}
\def\eeq{\begin{equation}}
\newcommand{\Aa}{\mbox{$\cal A\;$}}
\newcommand{\Dd}{\mbox{$\cal D\;$}}
\newcommand{\Hh}{\mbox{$\cal H\;$}}
\newcommand{\Ll}{\mbox{$\cal L\;$}}
\newcommand{\r}{\mbox{\bf r}}

\def\pmb#1{\setbox0=\hbox{$#1$}%
\kern-.025em\copy0\kern-\wd0
\kern.05em\copy0\kern-\wd0
\kern-.025em\raise.0433em\box0}

\skip\footins = 14pt % was 12
\begin{document}

\nonfrenchspacing
\flushbottom
\title{Gauge Theories in the Momentum/Curvature Representation}

\footnotetext[1] {\baselineskip=16pt This work is supported in part by funds
provided by  the U.S.~Department of Energy (D.O.E.) under contract
\#DE-FC02-94ER40818. \hfil MIT-CTP-2516 \hfil  March 1996\break}

\author{R.~Jackiw\footnotemark[1]}

\address{Center for Theoretical Physics\\ Massachusetts Institute of
Technology\\ Cambridge, MA ~02139--4307}

\maketitle

\setcounter{page}{0}
\thispagestyle{empty}

\vskip 5in

\centerline{American Mathematical Society, New York~NY, April 1996}

\newpage

\baselineskip=24pt plus 1pt minus 1pt

\section{Introduction}

I shall discuss some kinemetical properties of non-Abelian quantum gauge
theories that are elementary, but not widely appreciated.  Use will be made
of various nice mathematical structures, so I hope the material will interest
this audience.

Consider a generic non-abelian gauge theory, with action presented in first
order form.
\begin{equation}
I=\int dt \int d^d r \left[ E^i_a \dot{A}^a_i - \Hh(E,A)+A^a_0 G_a\right]
\label{eq:1}
\end{equation}
Here the covariant spatial components of the gauge potential (connection)
$A^a_i$ are the canonical coordinates and the conjugate momenta are identified
from the first term in the integrand~(\ref{eq:1}) as $E^i_a$.  
Indeed the first term is the (functional) canonical 1-form, analogous to the 
1-form arising in particle mechanics:$\int dt p\dot{q}=\int pdq$  (overdot
indicates time-differentiation). Further, in~(\ref{eq:1}) \Hh denotes the
Hamiltonian density, and $G_a$ is the Gauss-law generator whose vanishing is
enforced by the Lagrange multiplier $A_0^a$, which is also the temporal
component of the gauge potential.  The dimensionality $d$ of space, over whose
volume the spatial integral is taken, has been left unspecified; also unspecified
is the explicit form for \Hh.  However we assume that $G_a$ generates the usual
gauge transformation on $A_{i}^a$ with parameters $\theta^a$ and structure
constants $f_{bc} \; ^a = - f_{cb} \; ^a$,
\begin{equation}
\delta_\theta\,A^a_i = \partial_i  \theta^a + f_{bc} \;^a \,A_i^b\theta^c
\label{eq:2}
\end{equation}
Further, we assume that $E_a^i$ transforms covariantly.
\begin{equation}
\delta_\theta\,E_a^i = - f_{ab}\; ^c \, \theta^b\,E^i_c
\label{eq:3}
\end{equation}

Various gauge field models fit our requirements, but not theories with a
Chern-Simons term: for these the canonical momentum does not transform
covariantly.  The usual Yang-Mills model in any spatial dimension satisfies the
desired requirements.  Its dynamics, conventionally derived from the
second-order Lagrange density
\[
\Ll_{YM}  =  -\quar F^a_{\mu\nu} F^{\mu\nu} \; ^a 
\]
\vskip-.1in
\begin{mathletters}
\begin{equation}
F^a_{\mu\nu}  =  \partial_\mu A^a_\nu -\partial_\nu A^a_\mu + f_{bc}\;^a A^b_\mu 
A^c_\nu %\eqno{(4a)}
\label{eq:4a}
\end{equation}
can be equivalently encoded in the first order action~(\ref{eq:1}), with
\begin{equation}
\Hh = \half E^i_a E^i_a + \quar F^a_{ij} F^{ija} %\eqno{(4b)}
\label{eq:4b}
\end{equation}
and
\begin{equation}
G_a=(D_iE^i)_a = \partial_i E^i_a + f_{ab}\;^c \, A^b_i E^i_c %\eqno{(4c)}
\label{eq:4c}
\end{equation}
\end{mathletters}
where the canonical momentum $E^i_a$ coincides with the non-Abelian electric field
(curvature) $F^a_{oi}$. For $d=1$, there is no magnetic contribution to \Hh, since
$F^a_{ij}$ does not exist.  Additionally, in one spatial dimension, another gauge theory
may be constructed:  the so-called
``$B$-$F$'' model, described by the covariant Lagrange density
\begin{mathletters}
\begin{equation}
\Ll_{BF}=\half \eta_a\varepsilon^{\mu\nu} F^a_{\mu\nu} %\eqno{(5a)}
\label{eq:5a}
\end{equation}
that is already in first order form, and so is governed by the
action~(\ref{eq:1}), with vanishing \Hh, and
\begin{equation}
G_a=(D\eta)_a=\eta_a'+ f_{ab}\;^c \, A^b_1\eta_c %\eqno{(5b)}
\label{eq:5b}
\end{equation}
\end{mathletters}
(Dash indicates differentiation with respect to the single spatial coordinate $x$;
$\varepsilon^{\mu\nu}$ is the two-dimensional anti-symmetric tensor.) The gauge
covariant
$\eta_a$ is identified with
$E^1_a$.  (The
$B$ in the
``$B$-$F$'' nomenclature refers to $B_a^{\mu\nu} \equiv \half
\eta_a  \varepsilon^{\mu\nu}$.)  Note that owing to the vanishing of the ``$B$-$F$''
Hamiltonian, the problem of solving a ``$B$-$F$'' quantum theory reduces to solving its
Gauss Law, {\it i.e.}~finding states annihilated by $G_a$ of~(\ref{eq:5b}).

In the quantized theory, the canonical variables satisfy equal-time commutation
relations
\begin{equation}
i[E^i_a(\r), A^b_j(\r')] = \delta^i_j \delta^b_a
\delta(\r-\r') %\eqno{(6)}
\label{eq:6}
\end{equation}
and the gauge transformation rules~(\ref{eq:2}),~(\ref{eq:3}) are gotten by
commuting with
$\int d^d r \theta_a G^a \equiv G_\theta$.
\begin{mathletters}
\begin{equation}
[G_\theta, A^a_{i}]=i\delta_\theta A^a_i %\eqno{(7a)}
\label{eq:7a}
\end{equation}
\begin{equation}
[G_\theta,E^i_a]=i\delta_\theta E^i_a %\eqno{(7b)}
\label{eq:7b}
\end{equation}
\end{mathletters}
(A common time argument in all operators is suppressed.)

An explicit realization is given in a Schr\"odinger representation, where states
are described by wave functionals of $A,\Psi(A)$, and the action of the operator
$A$ is realized by multiplication by $A$, while $E$ is realized by functional
differentiation:  $E\sim\Ii\frac{\delta}{\delta A}$.  Moreover, physical states
are annihilated by $G_a$, which also means that the wave functionals are gauge
invariant
\begin{equation}
\Psi(A^U_i)\equiv \Psi(U^{-1} A_i U+U^{-1}\partial_i U)=\Psi(A_i) %\eqno{(8)}
\label{eq:8}
\end{equation}
(Frequently we use group-index free notation:  $A_i\equiv A^a_i T_a$, {\it etc.}, 
where
$T_a$ are anti-Hermitian Lie algebra generators; also $\langle A,E\rangle \equiv
A_i^a E^i_a$.) For $d=3$, where the gauge transformation $U$ can be homotopically
non-trivial, a phase involving the vacuum angle may arise in the response of the
wave functional to a gauge transformation; here I shall ignore this complication.

The realization described above is the field theoretical analog of the quantum
mechanical story, where  wave functions depend on $q,\psi(q)$, the operator $q$
acts by multiplication and $p$ is realized as a derivative $\Ii \frac{d}{dq}$.
But in quantum mechanics, we may also use the momentum
representation, where $p$ acts by multiplication on wave functions that depend on
$p,\varphi(p)$, and $q$ is  realized by differentiation $i\frac{d}{dp}$.  The
relation between the two is given by a Fourier transformation.
\begin{equation}
\varphi(p) = \int \frac{dq}{\sqrt{2\pi}} e^{-ipq} \psi(q) %\eqno{(9)}
\label{eq:9}
\end{equation}

I shall discuss here some properties of the field theoretic momentum
representation, where $E$ acts by multiplication on wave functionals that depend
on $E$, while $A$ is realized by (functional) differentiation as
$i\frac{\delta}{\delta E}$.$^1$

\section{Response to Gauge Transformations}

While physical states in the ``$A$'' representation are gauge invariant, see
Eq.~(\ref{eq:8}), those in the ``$E$'' representation are not.  This is immediately
established by using the (functional) Fourier transform relation between
functionals $\Phi(E)$ in the ``$E$'' representation, and the gauge invariant wave
functionals $\Psi (A)$ of the ``$A$'' representation.  The following chain of equations holds.
$$\Phi(E) = \int \Dd A\Big(\exp -i\int d^d r\langle E,A\rangle\Big)\Psi(A) $$
$$\hspace{1.5in} = \int \Dd A\Big(\exp -i\int d^d r\langle
E,A\rangle\Big)\Psi(U^{-1}
      AU+U^{-1}\partial U)$$
$$\hspace{1.1in} = \int \Dd A\Big(\exp -i\int d^dr\langle E^U,A\rangle\Big)\Psi 
(A + U^{-1}\partial U)$$
$$\hspace{2in} = \exp i\int d^d r \langle E, \partial U U^{-1}\rangle \int
      \Dd A \Big(\exp -i\int d^d r \langle E^U, A\rangle\Big)\Psi(A)$$
\begin{equation}
\!\!\!\!\!\!\!\!\!\!\!\!\!\!\!= \exp -i \Omega (E,U) \Phi (E^U) %\eqno{(10)}
\label{eq:10}
\end{equation}
The first equation is the field theoretic analog to~(\ref{eq:9}).  The second equality is
true because
$\Psi (A)$ is gauge invariant.  In the third equality we have changed integration
variables: 
$A\rightarrow UAU^{-1}$; this has unit Jacobian, and affects the phase by replacing
$E$ with its gauge transform
$E^U=U^{-1}EU$.  In the next step, $A_i$ is shifted:  $A_i\rightarrow A_i -
U^{-1}\partial_i U$; this produces the phase $\Omega(E,U)$ seen in the last
equality.
\begin{equation}
\Omega(E,U)=-\int d^d r E^i_a (\partial_i UU^{-1})^a %\eqno{(11)}
\label{eq:11}
\end{equation}

Thus from~(\ref{eq:10}), it follows that physical wave functionals in
the ``$E$'' representation are {\bf not} gauge invariant.  Rather, after a gauge
transformation they acquire the phase $\Omega(E,U),$
\begin{equation}
\Phi(E^U)=e^{i\Omega(E,U)}\Phi(E) %\eqno{(12)}
\label{eq:12}
\end{equation}
which is recognized to be a 1-cocycle {\it i.e.}~ $\Omega(E,U)$ satisfies
\begin{equation}
\Omega(E,U_1U_2)=\Omega(E^{U_1},U_2)+\Omega(E,U_1) %\eqno{(13)}
\label{eq:13}
\end{equation}
as is required by~(\ref{eq:12}) when two gauge transformations are composed.

We conclude therefore that physical functionals in the ``$E$'' representation, which
are annihilated by the Gauss law generator $G_a$, obey~(\ref{eq:12}).  Before
exploring further properties of that equation, let us give another perspective on
the result.\cite{ref:1}

\section{Geometric Quantization}

One may pose the following question:  why is it that functionals of the gauge
covariant variable $E$ are not gauge invariant, while functionals of the gauge
non-invariant variable $A$ are gauge invariant.  The answer lies in the fact that
the canonical 1-form $\int dt\int d^d r E^i_a \dot{A}^a_i$ is not gauge
invariant.  The best way to understand this statement is in the context of
geometric quantization.$^2$  So I shall first briefly review that formalism, using
for simplicity ordinary quantum mechanics as an illustration.

Collect the canonical variables $p,q$ into the pair $\xi^m: \xi^1=p,\xi^2=q$,
which serve as coordinates for the two-dimensional phase space.  The canonical
1-form $pdq$ is written as
$$
\theta=\theta_m d\xi^m
$$
\begin{equation}
\theta_1=0, \theta_2=p %\eqno{(14)}
\label{eq:14}
\end{equation}
while the symplectic 2-form reads
$$\omega = d\theta = \half \omega_{mn} d\xi^m d \xi^n$$
\begin{equation}
\omega_{mn}=\frac{\partial\theta_n}{\partial\xi^m} -
\frac{\partial\theta_m}{\partial\xi^n} = \varepsilon_{mn} %\eqno{(15)}
\label{eq:15}
\end{equation}
Canonical transformations are coordinate transformations on phase space that leave
$\omega$ invariant; infinitesimally they are given by a vector field $v^m(\xi)$.
\begin{equation}
\delta\xi^m=-v^m(\xi) %\eqno{(16)}
\label{eq:16}
\end{equation}
A `` generator'' $G(\xi)$ for a vector field $v^m$ is defined by
\begin{equation}
v^m \omega_{mn}=-\frac{\partial G}{\partial\xi^n} %\eqno{(17)}
\label{eq:17}
\end{equation}
Conversely, for any function $G(\xi)$ on phase space, we can use~(\ref{eq:17}) to
define a vector field $v^m$.  (It is assumed that $\omega_{mn}$ is
non-degenerate, {\it i.e.} it has an inverse $\omega^{mn}$; in our case
$\omega^{mn}=-\varepsilon^{mn}$.)

Within geometric quantization, there is a stage called ``pre-quantization'' that
arises before the conventional quantum theory is defined.  One works with
pre-quantized wave functions $f(\xi)$ that vary over the entire phase space,
{\it i.e.}~they depend on {\bf both} $p$ and $q$. To every quantity $G(\xi)$
one associates a pre-quantized operator $\widehat{G}$ that acts on the $f(\xi)$. 
The operator is given by
\begin{equation}
\widehat{G}=\Ii v^m \Dd\!\!_m + G %\eqno{(18)}
\label{eq:18}
\end{equation}
where $v^m$ is defined from $G$ by~(\ref{eq:17}), and $\Dd\!\!_m$ is the covariant
derivative
\begin{equation}
\Dd\!\!_m \equiv \frac{\partial}{\partial\xi^m} - i \theta_m %\eqno{(19)}
\label{eq:19}
\end{equation}

Thus the coordinate $q=\xi^2$ produces, according to~(\ref{eq:15}) 
and~(\ref{eq:17}), the vector field $v^m=(-1,0)$ and according to~(\ref{eq:14}),
~(\ref{eq:18}) and~(\ref{eq:19}),  the pre-quantized operator is
\begin{equation}
\hat{q}=\Ii v^m \Dd\!\!_m + q = i\frac{\partial}{\partial p} + \theta_1 + q =
i\frac{\partial}{\partial p} + q %\eqno{(20)}
\label{eq:20}
\end{equation}
Similarly, $p=\xi^1$ leads  to $v^m=(0,1)$, and
\begin{equation}
\hat{p}=\Ii v^m \Dd\!\!_m + p = \Ii \frac{\partial}{\partial q} - \theta_2 + p =
\Ii \frac{\partial}{\partial q}% \eqno{(21)}
\label{eq:21}
\end{equation}

Finally the quantum theory, with its Hilbert space, is defined by choosing a
``polarization''.  This consists of fixing polarization vector fields $\pi^m$,
which span {\bf half} the (even-dimensional) phase space, and imposing on the
pre-quantized functions $f(\xi)$ the conditions
\begin{equation}
\pi^m \Dd\!\!_m \,f=0 %\eqno{(22)}
\label{eq:22}
\end{equation}
Equations~(\ref{eq:22}) determine dependence on half the phase-space coordinates, leaving
arbitrary the dependence on the other half, and quantum mechanical wave functions
are solutions to~(\ref{eq:22}).

Selection of the conventional ``coordinate'' representation is accomplished by
using the vector field corresponding to $q=\xi^2:\pi^m=(-1,0)$.  Wave functions in
the coordinate polarization therefore satisfy
\begin{mathletters}
\begin{equation}
\Dd\!\!_1 f_{\hbox{\rm\tiny coordinate}} = \frac{\partial}{\partial p} f_{\hbox{\rm\tiny
coordinate}} = 0 %\eqno{(23a)}
\label{eq:23a}
\end{equation}
which is solved by arbitrary functions of $q$ that become the quantum mechanical
wave functions.
\begin{equation}
f_{\hbox{\rm\tiny coordinate}}=\psi(q) %\eqno{(23b)}
\label{eq:23b}
\end{equation}
\end{mathletters}
The operators $\hat{q}$ and $\hat{p}$, whose form is given in~(\ref{eq:20})
and~(\ref{eq:21}), act as expected.
$$\hat{q} f_{\hbox{\rm\tiny coordinate}} = q\psi(q)$$
\begin{equation}
\hat{p}f_{\hbox{\rm\tiny coordinate}}= \Ii \frac{d}{d q}\psi(q)
%\eqno{(24)}
\label{eq:24}
\end{equation}

The alternative ``momentum'' polarization uses the vector field corresponding to
$p=\xi^1:\pi^m =(0,1)$.  The polarization condition becomes
\begin{mathletters}
\begin{equation}
\Dd\!\!_2 f_{\hbox{\rm\tiny momentum}} = \left(\frac{\partial}{\partial q} - i p\right)
f_{\hbox{\rm\tiny momentum}} = 0 % \eqno{(25a)}
\label{eq:25a}
\end{equation}
which is solved by arbitrary functions of $p$, times a phase involving $q$
\begin{equation}
f_{\hbox{\rm\tiny momentum}} = e^{ipq} \varphi (p) %\eqno{(25b)}
\label{eq:25b}
\end{equation}
\end{mathletters}
For the quantum mechanical wave function, the phase is stripped away from the
pre-quantized function, leaving $\varphi(p)$.  Action of operators $\hat{q}$ and
$\hat{p}$ on $\varphi$ is deduced from their action on $f$.
\begin{mathletters}
\begin{equation}
\hat{q}f_{\hbox{\rm\tiny momentum}} = 
\left(i\frac{\partial}{\partial p} + q\right) 
e^{ipq} \varphi (p)=e^{ipq}i
\frac{d}{d p} \varphi(p)
%\eqno{(26a)}
\label{eq:26a}
\end{equation}
\begin{equation}
\hat{p} f_{\hbox{\rm\tiny momentum}} =
\Ii \frac{\partial}{\partial q} e^{ipq}\varphi(p) = e^{ipq}
p\varphi(p) %\eqno{(26b)}
\label{eq:26b}
\end{equation}
\end{mathletters}
Once again the expected formulas emerge in the action on $\varphi(p)$.

We are ready now to examine our gauge theory within the above formalism.  Since
the symplectic 1-form $\int dt \int d^d r E^i_a \dot{A}^a_i$ is the field theoretic
generalization of the particle expression $\int pdq$, we may immediately take over
the previous results, with field variables replacing particle variables 
$(p\rightarrow E, q\rightarrow A)$ in a pre-quantized wave functional depending on
$E$ and $A$, $F(E,A)$.

Next we determine the pre-quantized operator that corresponds to $G_\theta$.  A
straight forward calculation shows that
\begin{equation}
\widehat{G}_\theta = i\int d^d r\left( \delta_\theta E^i_a \frac{\delta}{\delta
E^i_a} + \delta_\theta A^a_i \frac{\delta}{\delta A^a_i} \right) %\eqno{(27)}
\label{eq:27}
\end{equation}
{\it i.e.} $\widehat{G}_\theta$ effects an infinitesimal gauge transformation on the
pre-quantized wave functional $F(E,A)$.  Moreover, demanding that
$\widehat{G}_\theta$ annihilate $F$, thereby imposing Gauss' law at the
pre-quantized level, ensures that $F$ is gauge invariant.

But now recall that in the coordinate polarization the pre-quantized wave
functional, restricted to depend solely on $A$, coincides with the quantized wave
functional $\Psi(A)$.
\begin{equation}
F_{\hbox{\rm\tiny coordinate}} = \Psi(A) %\eqno{(28)}
\label{eq:28}
\end{equation}
Therefore $\Psi(A)$ is per
force gauge invariant.  On the other hand, with momentum polarization the gauge
invariant pre-quantized functional is given by
\begin{equation}
F_{\hbox{\rm\tiny momentum}} = e^{i\int d^d r \langle E,A\rangle }\Phi (E)
%\eqno{(29)}
\label{eq:29}
\end{equation}
Hence gauge invariance of $F$ requires
\begin{equation}
e^{i\int d^d r\langle
E^U,A^U\rangle}\Phi(E^U)=e^{i\int d^d r\langle E,A\rangle}\Phi(E) %\eqno{(30)}
\label{eq:30}
\end{equation}
This then is equivalent to~(\ref{eq:12}).

To summarize:  in geometric quantization, the pre-quantized wave functional is
gauge invariant, and so is the quantum wave functional in the coordinate
polarization.  But in the momentum polarization, owing to the gauge non-invariance
of the canonical 1-form, the quantum wave functional is not gauge invariant.

\section{Properties of the Cocycle and Wave Functional}

From the gauge transformation law~(\ref{eq:12}) for the wave functional $\Phi$, we
can deduce some of $\Phi$'s properties.  Of course no gauge invariant portion of
$\Phi$ is affected by~(\ref{eq:12}); so no information will be forthcoming on this
aspect of the wave functional.

It must be emphasized that non-trivial information is available only in the
non-Abelian case.  For an Abelian theory, with gauge invariant $E$ and
$U=e^{i\theta}$, where $\theta$ is function (not a matrix), it follows 
from~(\ref{eq:11}) that $\Omega(E,U)=\int d^d
r\partial_i E^i\theta$ and~(\ref{eq:12}) or~(\ref{eq:30}) merely require that $\Phi$
have support only on the transverse part of $E^i$.  This is the momentum-space
analog of the position-space condition that $\Psi(A)$ in the Abelian theory has
support only on the transverse (gauge invariant) portion of $A_i$.

Returning now to the non-Abelian case, we extract a gauge non-invariant eikonal factor from
the wave functional, leaving a gauge invariant functional $\hat{\Phi}(E)$.
\begin{mathletters}
\begin{equation}
\Phi(E)=e^{iS(E)} \hat{\Phi}(E) %\eqno{(31a)}
\label{eq:31}
\end{equation}
\begin{equation}
\hat{\Phi}(E^U)=\hat{\Phi}(E) %\eqno{(31b)}
\label{eq:31b}
\end{equation}
\end{mathletters}%
Note that the gauge invariant functional $\hat{\Phi}(E)$ is annihilated by the ``rotation''
part of the Gauss generator $if_{ab}\; ^c \, E^i_c A^b_i \hat{\Phi}(E) = if_{ab}\; ^c \,
E_c^i {\delta \over
\delta E_b^i} \hat{\Phi}(E) = 0; \hat{\Phi}(E)$ is {\bf not} annihilated by the full Gauss
generator owing to its $\partial_i E^i_a$ part.  (Thus we see again that physical wave
functionals in the ``$E$'' representation cannot be gauge invariant, because gauge
invariant functionals are {\bf not} annihilated by the Gauss generator.) 
From~(\ref{eq:12}) and (31) it follows that $S(E)$ must satisfy 
\begin{equation}
S(E^U)-S(E)=-\int d^d r E^i_a(\partial_i UU^{-1})^a %\eqno{(32)}
\label{eq:32}
\end{equation}
[An integer multiple of $(2\pi)$ can also be present in~(\ref{eq:32}).]  This formula
would indicate that the 1-cocycle is trivial, since it appears expressible as a
coboundary, {\it i.e.}~as the difference on the left side of~(\ref{eq:32}).  However,
such a conclusion would be misleading because $S(E)$ is necessarily singular.  To
see that, present~(\ref{eq:32}) infinitesimally as
\begin{equation}
S(E+[E,\theta])-S(E) = \int d^d r\partial_i E^i_a \theta^a  %\eqno{(33)}
\label{eq:33}
\end{equation}
If $S$ is a non-singular functional of $E^i$, we can choose $E^i$ so
that it commutes with $\theta$ in the Lie algebra, whereupon the left side
vanishes, while the right side need not.  The contradiction is resolved by
allowing $S(E)$ to possess singularities, see below.

Before attempting to solve for $S(E)$ from~(\ref{eq:32}), let us observe that
~(\ref{eq:32}) also implies that the quantity
\begin{equation}
\Aa\!\!^a_i (E) \equiv - \frac{\delta S(E)}{\delta E^i_a} %\eqno{(34)}
\label{eq:34}
\end{equation}
transforms as a gauge connection
\begin{equation}
\Aa\!\!_i (E^U)=U^{-1}\Aa\!\!_i(E)U+U^{-1}\partial_i U %\eqno{(35)}
\label{eq:35}
\end{equation}
This suggests that a formula for $S(E)$ could have the form
\begin{equation}
S(E)=-\int d^d r  E^i_a(g^{-1}\partial_i g)^a %\eqno{(36)}
\label{eq:36}
\end{equation}
where $g$ is an as-yet-to-be-determined functional of $E$, with the property that
it transforms like a group element.
\begin{equation}
g(E^U)=g(E)U %\eqno{(37)}
\label{eq:37}
\end{equation}
The gauge connection~(\ref{eq:34}) becomes
\begin{equation}
\Aa\!\!^a_i = (g^{-1}\partial_i g)^a - \int d^d r\partial_j(gE^j g^{-1})_b \left(
\frac{\delta g}{\delta E^i_a} g^{-1}\right)^b% \eqno{(38)}
\label{eq:38}
\end{equation}

We now discuss separately the one-dimensional $(d=1)$ and the higher-dimensional
$(d\geq~2)$ models.  

\subsection{One Dimension}

In one dimension, the Gauss law reads
\begin{equation}
\left(\eta'_a + f_{ab}\; ^c \,  \eta_c i\frac{\delta}{\delta\eta_b}\right) \Phi
(\eta) = 0 %\eqno{(39)}
\label{eq:39}
\end{equation}
(We have replaced $E^1_a$ by $\eta_a$.)
Contracting this equation with $\eta_a$ and using anti-symmetry of the structure
constants to eliminate the second term shows that $\Phi(\eta)$ has support only on
vanishing $(\eta_a\eta_a)'$, {\it i.e.}~on $\eta$ fields that are in the orbit of a
constant.  Consequently in~(\ref{eq:38}) we can choose $g$ to be that group element which
takes $\eta$ to the constant, so that the last term is absent.
$$
\eta = g^{-1}Kg
$$
\begin{equation}
K \;{\hbox{\rm constant and invariant}}% \eqno{(40)}
\label{eq:40}
\end{equation}
It then follows that $S$ may be written as
\begin{equation}
S(\eta) = - \int dx K_a \left(\frac{dg}{dx}g^{-1} \right)^a
%\eqno{(41)}
\label{eq:41}
\end{equation}
with $g$ related to $\eta$ by~(\ref{eq:40}).  The transformation law~(\ref{eq:32}) is
straightforwardly verified from~(\ref{eq:37}) and~(\ref{eq:40}).  Note that the 
connection~(\ref{eq:34}) becomes a pure gauge.  \cite{ref:4},\cite{ref:5}

The above structure~(\ref{eq:41}) has another role in mathematical physics, quite distinct
from the role in which we encounter  it here as the phase of a wave functional. 
Observe that $S$ in~(\ref{eq:41}) is given by an integral of the 1-form $\langle
K,dgg^{-1}\rangle$, which one may take as a canonical 1-form for a Lagrangian
with variables depending on ``time''.  It is then further true that the
symplectic 2-form, $d\langle K,dgg^{-1}\rangle =\langle K,dgg^{-1}
dgg^{-1}\rangle$ defines Poisson brackets and that the brackets of the quantities
$Q^a = (g^{-1} Kg)^a$ reproduce the Lie algebra of the relevant group.   This
2-form is associated with the names Kirillov and Kostant.$^3$  [One recognizes here
a development that has previously occurred in connection with the Chern-Simons
term:  this term first arose in physics as the phase of the QCD wave functional in
(3+1)-dimensional Yang-Mills theory, whose gauge transformation response gives rise to the
vacuum angle.  Subsequently the Chern-Simons term was used in dynamics for a
lower-dimensional field theory.$^4$]

Formula~(\ref{eq:41}) can be presented for any group, but it is not explicit, in that the
group element's dependence on $\eta$ is defined only implicitly by~(\ref{eq:40}).  For a
specific Lie group, an explicit formula may be given by expressing $g$ in terms of
$\eta$.  For example for $SU(2)$ one finds
$$
S(\eta) = \six\int dx \varepsilon^{abc}\eta'_a\tan^{-1}\frac{\eta_b}{\eta_c}$$
\begin{equation}
\eta_a\eta_a \;{\hbox{\rm constant}} %\eqno{(42)}
\label{eq:42}
\end{equation}
This expression also puts into evidence the singularities of $S$ that were
mentioned earlier.  Finally we remark that since the ``$B$-$F$'' Hamiltonian vanishes,
the entire problem of quantization, which reduces to satisfying the Gauss law, is
solved by the wave functionals~(\ref{eq:31}), with $S$ given by~(\ref{eq:41}).

\subsection{Higher Dimensions}

In higher dimensions there does not appear to be a general, mathematically elegant,
formula for $S(E)$ valid for arbitrary groups.  Specific expressions can be given,
and for $SU(2)$ one has \cite{ref:3}
\begin{equation}
S(E)=\frac{1}{d-1}\int d^d r \varepsilon^{abc}(E^i_a E^j_b \partial_i E^k_c)
\phi_{jk}% \eqno{(43)}
\label{eq:43}
\end{equation}
where the gauge invariant $\phi_{jk}$ is defined by
\begin{equation}
E^i_a E^j_a \phi_{jk} = \delta^i_k %\eqno{(44)}
\label{eq:44}
\end{equation}
and exhibits the singularities that necessarily are present in~(\ref{eq:43}). 
 With~(\ref{eq:43}),
the connection~(\ref{eq:34}) is no longer a pure gauge; it gives rise to non-vanishing
curvature.

\section{Applications}

The above ideas may have application in analyzing gauge theories -- the Yang-Mills
models in spatial dimension greater than one, as well as ``$B$-$F$'' theories, which
are defined on a line and have been used to model lineal gravity.

\subsection{Yang-Mills Theories}

In the Yang-Mills case, the above line of research is motivated by the expectation
that the mysteries of non-Abelian gauge theories at low energy -- like, for
example, confinement and the spectrum of low-lying states -- can be unraveled when
gauge covariant variables -- like $E$ -- are used.  In this connection, the
singularities of $S(E)$ are viewed as analogous to the centrifugal barrier that is
seen in particle quantum mechanics, when radial (rotation covariant) variables
are used.  It is hoped that analysis of these singularities will provide clues to
low energy dynamics -- but it is also true that thus far the hope has not been
fulfilled.

\subsection{Gravity Theories}

In the one-dimensional case, the ``$B$-$F$'' theory arises in gauge theoretic
reformulations of lineal gravity.  This comes about in the following fashion.

If one wants to construct a gravitational theory on a line, {\it i.e.}~in
(1+1)-dimensional space-time, Einstein's general relativity model cannot be
used, because the Einstein tensor (which enters the general relativistic field
equation) vanishes identically, since in two dimensions, the Ricci tensor
$R_{\mu\nu}$ is proportional to the scalar $R: R_{\mu\nu} -\half g_{\mu\nu} R=0$. 
Correspondingly the Einstein-Hilbert Lagrange density $\sqrt{-g}R$ is a total
derivative, and does not give rise to Euler-Lagrange equations.

A way around this impasse was suggested some years ago.  One introduces an
additional non-geometrical, world scalar variable $\eta$ and uses, instead of the
Einstein-Hilbert formula, the Lagrange density \cite{ref:9}
\begin{equation}
\Ll = \sqrt{-g} \eta R +\ldots %\eqno{(45)}
\label{eq:45}
\end{equation}
where the ellipsis stands for further $\eta$-and metric-dependent terms, which
give rise to different theories.  More recently models of this type have been
abstracted from string theory, and in this context they are called
``dilaton-gravity'' theories, with $ln\eta$ being identified as the ``dilaton''
field.  Alternatively, they are also known as ``scalar-tensor theories'', $\eta$
being the scalar and $g_{\mu\nu}$ the tensor.

It turns out that several specific scalar-tensor models, with specific expressions
for the ellipsis in the above Lagrange density, can be equivalently formulated as
gauge theories of the ``$B$-$F$'' variety.  Such formulations proceed along the
following steps. \cite{ref:10},\cite{ref:11}

Step 1.  For gravitational variables do {\bf not} use the metric tensor, but rather
the Einstein-Cartan variables:  the {\it Vielbein} $e^a_\mu$ and the
spin-connection $\omega^{ab}_{\mu}$.  Here, as before, Greek letters index
space-time components, while Roman letters denote components in a flat tangent space,
with metric $\eta_{ab}$. [Note that  the present tangent space index ``a'' does not
have the same meaning as in Sections I-IV, where it ranged over the Lie algebra.]
The metric tensor is given by
\begin{equation}
g_{\mu\nu}=e^a_\mu e^b_\nu \eta_{ab} %\eqno{(46)}
\label{eq:46}
\end{equation}
In the two-dimensional application, we may set $\omega^{ab}_\mu =
\varepsilon^{ab}\omega_\mu$, and $\eta_{ab}=\,\hbox{\rm diag}\, (1,-1)$.  In
addition to $e^a_\mu$ and $\omega_\mu$, it may be necessary to use further
variables, see below.  At this stage one has in hand a gauge theory of the local
Lorentz group, which in two space-time dimensions contains the single
generator $J$, and $\omega_\mu$ is the associated gauge potential.  The
{\it Zweibein}
$e^a_\mu$ transforms covariantly under the Lorentz group -- it is not a potential.

Step 2.  To have a completely gauge theoretic description of the gravity theory,
we consider translations, generated by $P_a$, and take the {\it Zweibeine} to be
the associated gauge potentials. 

Step 3.  To close the algebraic system, we look to the Lie algebra of the
generators $J$ and $P_a$.  As is conventional, we let $J$ generate rotations on
$P_a$
\begin{equation}
[P_a,J] = \varepsilon_a \; ^b \, P_b %\eqno{(47)}
\label{eq:47}
\end{equation}
But for the $[P_a,P_b]$ commutator we have a choice:  (1)~it can vanish -- as in
the three-parameter Poincar\'{e} group; (2)~it can be proportional to $J$ -- as in the
three-parameter DeSitter or anti-DeSitter groups; (3)~it can close on a central
element that commutes with $P_a$ and $J$ -- as in the centrally extended Poincar\'{e}
group, and this option is available only in two dimensions.  So we take for the
first two choices
\begin{mathletters}
\begin{equation} 
[P_a,P_b]=\varepsilon_{ab} \lambda J %\eqno{(48a)}
\label{eq:48a}
\end{equation}
or for the third choice
\begin{equation}
[P_a,P_b]=\varepsilon_{ab} I %\eqno{(48b)}
\label{eq:48b}
\end{equation}
\end{mathletters}
In Eq.~(\ref{eq:48a}), choice (1) above is realized in the limit $\lambda\rightarrow 0$. 
In Eq.~(\ref{eq:48b}), $I$ is the central element.  That quantity is taken as an additional
generator, commuting with $J$ and $P_a$, so the centrally extended Poincar\'{e}
group is viewed as a four-parameter group.  Consequently, if Eq.~(\ref{eq:48b}) is chosen, a
further gauge potential must supplement $\omega_\mu$ and $e^a_\mu$, we call it
$a_\mu$ and associate it with $I$.

Step 4.  A Lie-algebra valued connection is constructed as
\begin{equation}
A_\mu=e^a_\mu P_a + \omega_\mu J+a_\mu I %\eqno{(49)}
\label{eq:49}
\end{equation}
The curvature is constructed by the usual formula
\begin{mathletters}
\begin{equation}
F_{\mu\nu} = \partial_\mu A_\nu - \partial_\nu A_\mu + [A_\mu,A_\nu]% \eqno{(50a)}
\label{eq:50a}
\end{equation}
where the commutator is evaluated from~(\ref{eq:47}) and~(48).  Eq.~(\ref{eq:50a})
defines the curvature components
\begin{equation}
F_{\mu\nu}=f_{\mu\nu}^a P_a + f_{\mu\nu} J + a_{\mu\nu} I %\eqno{(50b)}
\label{eq:50b}
\end{equation}
\end{mathletters}
which can then be used in a ``$B$-$F$'' Lagrange density
\begin{equation}
\Ll_{BF} =
\frac{\varepsilon^{\mu\nu}}{2} \left(\eta_a f_{\mu\nu}^a + \eta_2 f_{\mu\nu} +
\eta_3 a_{\mu\nu}\right)% \eqno{(51)}
\label{eq:51}
\end{equation}
The last term in~(\ref{eq:49}),~(\ref{eq:50b}) and~(\ref{eq:51}), refering to the central direction in the
Lie algebra,  is present only when~(\ref{eq:48a}) is used.  In this way the Lagrange density
of~(\ref{eq:5a}) arises in the description of lineal gravity.  [Note that the index
``$a$'' in the previous Sections, in particular in~(\ref{eq:5a}),  ranges over the entire Lie
algebra, while in the present sub-Section it denotes the two tangent space
components,
$a=\{0,1\}$].

The dynamics based on~(\ref{eq:51}) entails the
requirement that the covariant derivative of the Lagrange multiplier multiplet
$(\eta_a,\eta_2,\eta_3)$ vanishes (this is obtained by varying $A_\mu$), and
the condition that the curvature $F_{\mu\nu}$ vanishes (this is obtained by varying the
Lagrange multipliers.).

One then shows that if the chosen Lie algebra is~(\ref{eq:48a}), the above dynamics is
equivalent to that of the first-posited scalar-tensor gravity theory, which is
governed by
\begin{equation}
\Ll\!_1=\sqrt{-g}\eta(R-\lambda) %\eqno{(52)}
\label{eq:52}
\end{equation}
where $\sqrt{-g}R$ coincides with $2\varepsilon^{\mu\nu}\partial_\mu\omega_\nu$,
while the $\eta$ in~(\ref{eq:52}) coincides with $\eta_2$ in~(\ref{eq:51}).
\cite{ref:9},\cite{ref:10}

On the other hand if Lie algebra is chosen to be~(\ref{eq:48b}), the gauge theoretical
dynamics becomes equivalent to that of the recently much discussed, string-inspired
model, with Lagrange density
\begin{equation}
\Ll_2=\sqrt{-g}(\eta R-\lambda) %\eqno{(53)}
\label{eq:53}
\end{equation}
Note that in the gauge theoretical formulation based on the extended Poincar\'e
group, the  ``cosmological constant'' parameter $\lambda$ does not appear in~(\ref{eq:51}), even
though it is present in~(\ref{eq:53}).  In fact $\lambda$ arises as a solution to the gauge theoretic
dynamics: one finds
$\eta_3=\lambda$, while $\eta_2$ continues to be identified with  
 $\eta$ in~(\ref{eq:53}), and
$\sqrt{-g}R$ remains $2\varepsilon^{\mu\nu}\partial_\mu\omega_\nu$.
\cite{ref:11},\cite{ref:12}

Since quantization of the ``$B$-$F$'' gauge theory consists merely of solving its
Gauss law, and this has been accomplished by formulas~(\ref{eq:31}) and~(\ref{eq:41}), we conclude
that quantization of the above two gravity theories~(\ref{eq:52}) and~(\ref{eq:53}) can also be
completely and explicitly carried out.

Studying the quantum theory of these diffeomorphism invariant models, and also of the
more complicated models, where matter degrees of freedom are coupled to the gravity
variables, promises to teach us valuable lessons about the nature of quantum
gravity, albeit in the unphysical setting of lower dimensionality.  The lower
dimensionality precludes the existence of gravitons and their concomitant
non-renormalizable interactions.   But one retains the possibility of examining
other questions about quantum gravity:  the issue of quantizing a diffeomorphism
invariant theory, the problem of time, the nature of the Wheeler-DeWitt equation, {\it
etc.}
\cite{ref:13}
\newpage
\begin{center}{\large NOTES} \end{center}
\begin{enumerate}
\item[1.] The momentum/curvature representation for gauge theories was introduced by Goldstone and Jackiw,
and worked out in detail for $SU(2)$ in Ref.~[1].  Generalization to other groups was given by Faddeev {\it
et al.}, as well as Baluni and Grossman, Ref.~[2].  Recent work on  Yang-Mills theory in the curvature
representation is by Freedman {\it et al.}, Ref.~[3], while ``$B$-$F$'' theories, which arise in descriptions
of gravity on a line, are discussed by Cangemi and Jackiw, Ref.~[4], Amati {\it et al.}, Ref.~[5] and Strobl
{\it et al.}, Ref.~[6].
\item[2.] Derivation of equation~(\ref{eq:12}) within geometric quantization is due to V.~P.~Nair
(unpublished).
\item[3.] An elementary discussion, together with references to the mathematical
literature is in  Bak {\it et al.}, Ref.~[7].
\item[4.] For a discussion, see Jackiw in Ref.~[8].

\end{enumerate}

\end{document}